\documentclass[aps,prc,dlspace,twocolumn,a4paper]{revtex4}
\usepackage{amsfonts}
\usepackage{epsfig}
\usepackage{color}
\usepackage{graphicx}
\usepackage{amsmath}
\usepackage{float}
\newcommand{\beq}{\begin{equation}}
\newcommand{\eeq}{\end{equation}}
\newcommand{\bqa}{\begin{eqnarray}}
\newcommand{\eqa}{\end{eqnarray}}

\newcommand{\eps}{\epsilon_s}

\newcommand{\jpsi}{J / \psi}

\newcommand{\pt}{p_{{}_T}}

\newcommand{\spt}{S(p_{{}_T})}
\newcommand{\sinpt}{ \langle S(p_{{}_T})\rangle}
\newcommand{\sincl}{\langle S^{{}^{\mbox{incl}}} \rangle}
\newcommand{\sdir}{\langle S^{{}^{\mbox{dir}}} \rangle}

\newcommand{\msbar}{\overline{\mbox{\rm MS}}}

\newcommand{\lambdamsbar}{\Lambda_{\overline{\rm MS}}}
\begin{document}
\title{Bottomonium suppression in nucleus-nucleus collisions using effective fugacity quasi-particle model}
 \author{Indrani Nilima $^{a}$}
\email{nilima.ism@gmail.com}
\author{Vineet Kumar Agotiya$^{a}$}
\email{agotiya81@gmail.com}
 \affiliation{$^a$Centre for Applied Physics, Central University of Jharkhand
 Ranchi, India, 835 205}

\begin{abstract}
In the present article, we have studied the equation of state and dissociation temperature of bottomonium state by correcting the full Cornell
potential in isotropic medium by employing the effective fugacity quasi-particle Debye mass. We had also calculated the bottomonium suppression in an expanding,
dissipative strongly interacting QGP medium produced in relativistic heavy-ion collisions. Finally we compared our results with experimental data from
RHIC 200GeV/nucleon Au-Au collisions, LHC 2.76 TeV/nucleon Pb-Pb, and LHC 5.02 TeV/nucleon Pb-Pb collisions as a function 
of number of participants.

\end{abstract}
\maketitle
\noindent{\bf KEYWORDS}: Equation of State, Strongly Coupled Plasma, Heavy 
Quark Potential, String Tension, Dissociation Temperature, Quasi-particle debye mass

\noindent{\bf PACS numbers}: 25.75.-q; 24.85.+p; 12.38.Mh
; 12.38.Gc, 05.70.Ce, 25.75.+r, 52.25.Kn \\

\section{Introduction}
At the Relativistic Heavy-Ion Collider (RHIC) situated at Brookhaven National Laboratory (BNL) heavy-ion collisions have been studied. After the pioneer work done in 
the direction of suppression by Matsui and Satz, and some other development of the
potential models, suppression was observed by both SPS and RHIC~\cite{Matsui:86}. Due to the Debye screening of the Quantum Chromo-Dynamic (QCD) potential between the 
two heavy quarks, quarkonia suppression was originally claimed to be an unambiguous signal of the formation of a Quark-Gluon Plasma (QGP).
Quarkonia suppression was suggested to be a signature of the QGP and we can measure the suppression ($\Upsilon$ as well as $J/\psi$), both 
at RHIC and at the LHC. 

In heavy-ion collisions to determine the properties of the medium formed in A+A collisions and p + p collisions and whether the A+A collision deviates from simple 
superposition of independent p + p collisions. This deviation is quantified with the nuclear modification factor ($R_{AA}$). This factor is the ratio of the yield in 
heavy-ion collisions over the yield in p + p collisions, scaled by a model of the nuclear geometery of the collision. The value of $R_{AA}$=$1$ indicates no modification 
due to the medium. We can say that the probe of interest is suppressed in heavy-ion collisions if $R_{AA}$ is less than $1$. A quarkonia meson that forms on the 
outside surface will not dissociate regardless of the temperature of the medium because it doesn’t have a chance to interact with it. This is why we never see 
a $R_{AA}$ that is equal to zero. The suppression can also be affected by the QGP, the formation time of the quarkonia meson and the QGP lifetime as well. 
For instance, a high $\pt$ quarkonia meson could have a formation time long enough that it actually does not see the QGP at all and thus isn’t suppressed.

In the early days most of the interests were focused on the suppression of charmonium states \cite{Matsui:86,Karsch:1987pv} of collider experiments at SPS and RHIC,
but several observations are yet to be understood {\em namely} the suppression of $\psi$ (1S) does not increase from SPS to RHIC,
even though the centre-of-mass energy is increased by fifteen times. The heavy-ion program at the LHC 
may resolve those puzzles because the beam energy and luminosity are increased by ten times of that
of the RHIC. Moreover the CMS detector has excellent capabilities for muon detection and provides measurements of $\psi$(2S) and the 
$\Upsilon$ family, which enables the quantitative analysis of quarkonia. That is why the interest may be shifted to the bottomonium
states at the LHC energy.

A potential model for the phenomenological descriptions of heavy quarkonium suppression would be quite useful inspite of the progress of direct lattice QCD based
determinations of the potential.The large mass of heavy quaks and its small relative velocity, makes the use of non-relativistic quantum mechanics 
justifiable to describe the quarkonia in the potential models.This is one of the main goal of this present study and argue for the modification of the full Cornell 
potential as an appropriate potential for heavy quarkonium at finite temperature.QGP created at RHIC have a very low viscosity to entropy ratio i.e. 
$\eta/{\mathcal S} \ge 1/4\pi$\cite{star,vis1,shur,son} and in the non-perturbative domain of QCD, temperature close to  $T_c$ the quark matter in the QGP 
phase is strongly interacting.

In the present paper, we shall employ quasi-particle model for hot QCD equations of state~\cite{chandra1,chandra2} to extract the debye mass~\cite{vcandra} which
is obtained in terms of quasi-particle degrees of freedom. We first obtained the medium modified heavy quark potential in isotropic medium and estimate the 
dissociation temperature. Here, we have used the viscous hydrodynamics to define the dynamics of 
the system created in the heavy ion collisions. We have included only the shear viscosity and not included the bulk viscosity. We will look the issue of bulk 
viscosity in near future. 

Our work is organized as follows. In Sec.II., we briefly discuss our recent work on medium modified potential in isotropic medium. In the subsections II (a) and (b) 
we study the real and imaginary part of the potential in the isotropic medium and Effective fugacity quasi-particle model(EQPM) in subsection (c).
In section III we studied about binding energy and dissociation temperature of $\Upsilon$, $\Upsilon'$ and $\chi_b$ state considering isotropic medium.
 Using this effective potential and by incorporating quasi-particle debye mass, we have then developed the equation of state for strongly interacting matter and have
 shown our results on pressure,energy density and speed of sound etc. along with the lattice data . In Sec.IV, we have employed the aforesaid equation of state
to study the suppression of bottomonium in the presence of viscous forces and estimate the survival probability in a longitudinally expanding QGP. 
Results and discussion will be presented in Sec.V and finally, we conclude in Sec.VI.


\section{Medium modified effective potential in isotropic medium}

We can obtain the medium-modification to the vacuum potential by correcting its both Coulombic and string part with a dielectric
function $ \epsilon(p)$ encoding the effect of deconfinement~\cite{prc-vineet}
\begin{eqnarray}
\label{defn}
V(r,T)&=&\int \frac{d^3\mathbf p}{{(2\pi)}^{3/2}}
 (e^{i\mathbf{p} \cdot \mathbf{r}}-1)~\frac{V(p)}{\epsilon(p)} ~,
 \label{cornell_pot}
\end{eqnarray}

Here the functions, $\epsilon(p)$ and $V(p)$ are the Fourier transform (FT) of the dielectric permittivity and Cornell potential respectively. 
After assuming $r$- as distribution ($r \rightarrow$ $r \exp(-\gamma r))$ we evaluated the Fourier transform of the linear part $\sigma r\exp{(-\gamma r)}$ as
\begin{eqnarray}
\label{eq-6-3}
-\frac{i}{p\sqrt{2\pi}}\left( \frac{2}{(\gamma-i p)^3}-\frac{2}{(\gamma+ip)^3}
\right).
\end{eqnarray}
While putting $\gamma=0$, we can write the FT of the linear term $\sigma r$ as,
\begin{equation}
\label{eq-6-4}
\tilde{(\sigma r)}=-\frac{4\sigma}{p^4\sqrt{2\pi}}.
\end{equation}
Thus the FT of the full Cornell potential becomes
\begin{equation}
\label{vk}
{V}(p)=-\sqrt{(2/\pi)} \frac{\alpha}{p^2}-\frac{4\sigma}{\sqrt{2 \pi} p^4}.
\end{equation}

To obtain the real and imaginary parts of the potential, we put the temporal component of real and imaginary part in terms of retarded (or advanced) and symmetric 
parts in the Fourier space in isotropic medium which finally gives,
\begin{eqnarray}
Re D^{00}_{11}(\omega,p)= \frac{1}{2}\left( D^{00}_{R}+D^{00}_{A}\right)\nonumber\\
\label{R}~~ {\rm{and}}~~
Im D^{00}_{11}(\omega,p)= \frac{1}{2} D^{00}_{F}.
\label{F}
\end{eqnarray}

Let us now discuss, the real and imaginary part of the potential modified using the above define $Re D^{00}_{11}(\omega,p)$ and $Im D^{00}_{11}(\omega,p)$ 
along with Effective fugacity quasi-particle model (EQPM) in the next sub-sections.
\begin{figure*}
\vspace{-3mm}
\includegraphics[scale=.50]{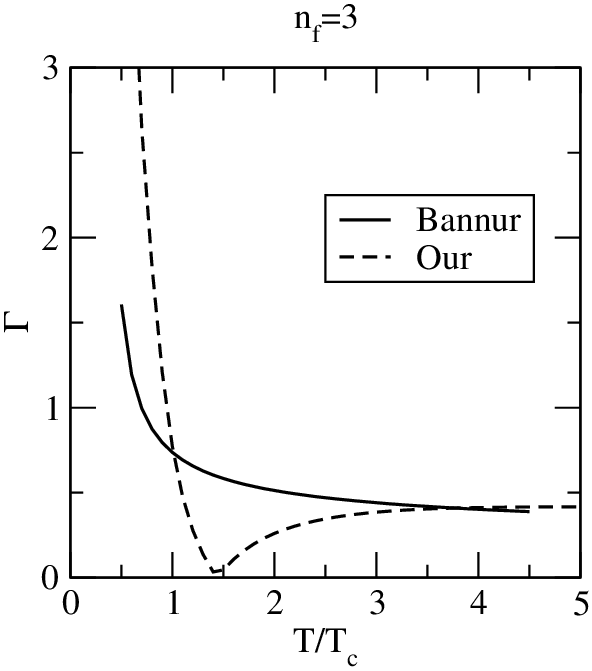}
\hspace{9mm}
\includegraphics[scale=.50]{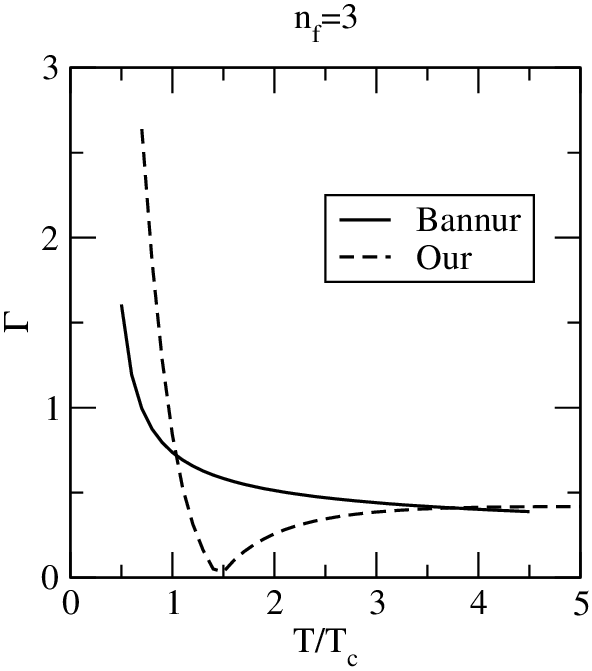}
\vspace{-1mm}
\caption{Plots of $ \Gamma $ as a function of $T/T_c$ for 3flavor QGP (extreme left figure) for EOS1~\cite{zhai,arnold} and for EOS2~\cite{kajantie} (extreme right figure).
In each figure, solid line represents the results obtained from Bannur EoS, dashed line represents
the results from our EoS (using quasi-particle Debye mass).}
\vspace{45mm}
\label{gama1}
\end{figure*}
\subsection{Real part of the potential in the isotropic medium}
Now using the real part of retarded (advanced) propagator in isotropic medium we get
\begin{equation}
Re D^{00}_{R,A}(0,p)=-\frac{1}{(p^2+m_D^2)}~,
\label{real_pot}
\end{equation}
whearas the real-part of the dielectric permittivity (also given in \cite{Schneider:prd66,Weldon:1982,Kapusta}) becomes
\begin{equation}
\epsilon (p)=\left( 1 + \frac{m_D^2}{p^2} \right)~.
\label{dielectriciso}
\end{equation}
 Now using Eq.\ref{real_pot} and real part of dielectric permittivity Eq.\ref{dielectriciso} in Eq.\ref{cornell_pot} we get,
 \begin{eqnarray}
\label{repot}
Re V_{(iso)}(r,T)&=&\int \frac{d^3\mathbf p}{{(2\pi)}^{3/2}} (e^{i\mathbf{p} \cdot \mathbf{r}}-1)\left(-\sqrt{(2/\pi)} \frac{\alpha}{p^2}
-\frac{4\sigma}{\sqrt{2 \pi} p^4}\right)\nonumber\\
&&\times\left( \frac{p^2}{(p^2+m_D^2)}\right) \nonumber\\
\end{eqnarray}
Solving the above integral, we find 
 \begin{eqnarray}
\label{pis}
Re V_{(iso)}(\hat{s},T)&=&\left(\frac{2\sigma}{m_D}-\alpha m_D\right)\frac{e^{-\hat s}}{\hat s}-\frac{2\sigma}{\hat{s}}\nonumber\\&&
+\frac{2\sigma}{m_D}-\alpha m_D~,
\end{eqnarray}
where $\hat{s} = r m_D$. In the limit $\hat{s}\ll1$, we have
\begin{eqnarray}
\label{lrp}
{Re V_{(iso)}(\hat{s},T)} \approx -\frac{2\sigma}{m{{}_D}\hat s}
-\alpha m_{{}_D},
\end{eqnarray}
\begin{figure*}
\vspace{-3mm}
\includegraphics[scale=.50]{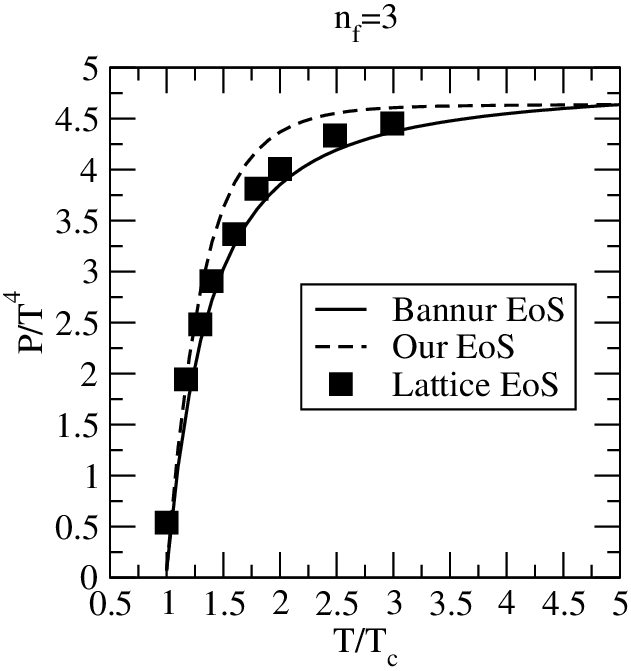}
\hspace{7mm}
\includegraphics[scale=.50]{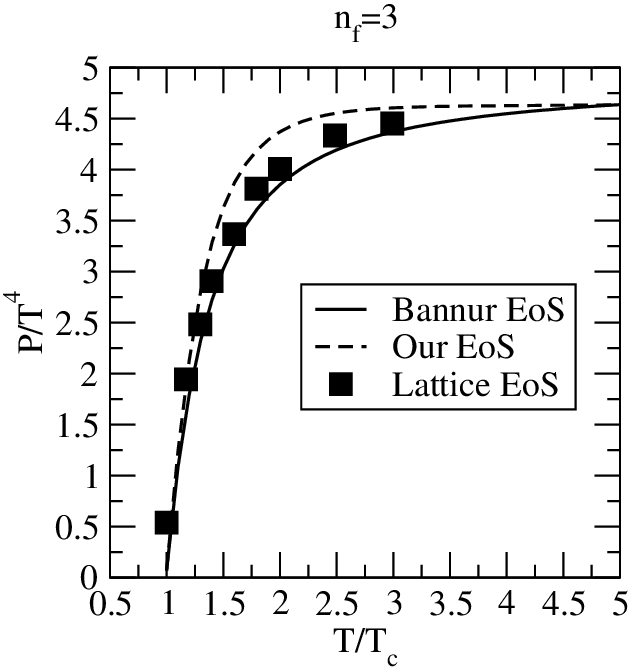}
\vspace{-1mm}
\caption{Plots of $ P/T^4 $ as a function of $T/T_c$ for 3-flavor QGP (extreme 
left figure) for EOS1~\cite{zhai,arnold} and for EOS2~\cite{kajantie} (extreme right figure). In each figure, solid line represents the results obtained 
from Bannur EoS, dashed line represents the results from Our EoS and 
diamond symbols represent lattice results~\cite{banscqgp, boyd}.}
\vspace{42mm}
\label{pres5}
\end{figure*}

\begin{figure*}
\vspace{-1mm}
\includegraphics[scale=.55]{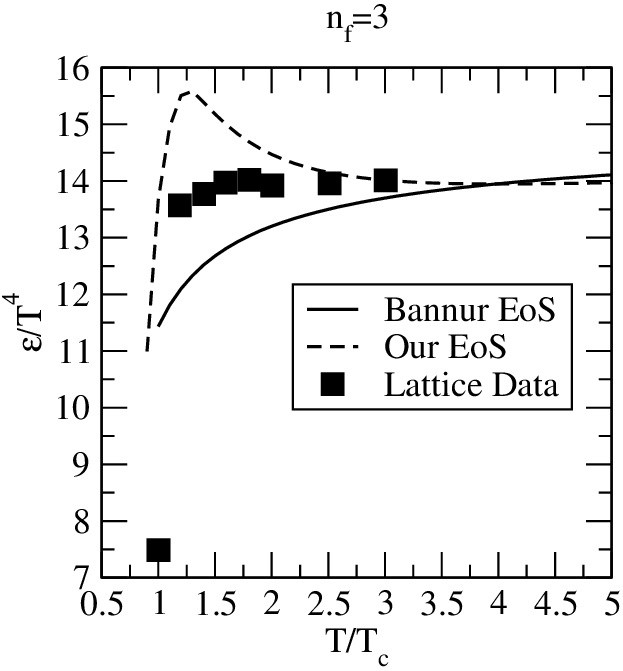}
\hspace{7mm}
\includegraphics[scale=.55]{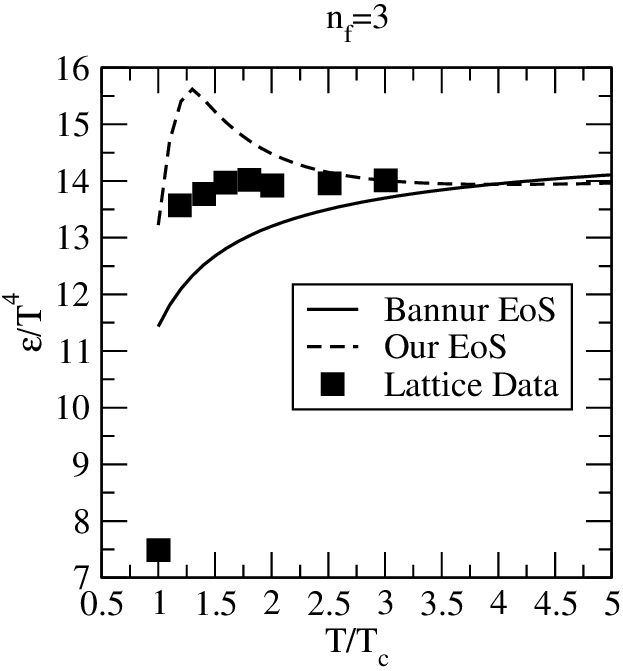}
\vspace{-1mm}
\caption{Plots of $\varepsilon/ T^4 $ as a function of $T/T_c$ for 
Our EoS (using quasi-particle Debye mass) and lattice results~\cite{banscqgp, boyd} for 3-flavor QGP (extreme left figure) for EoS1~\cite{zhai,arnold},
and for EOS2~\cite{kajantie} (extreme right figure). The notations are same as Figure2.}
\vspace{40mm}
\label{eps2}
\end{figure*}

\begin{figure*}
\vspace{-1mm}
\includegraphics[scale=.55]{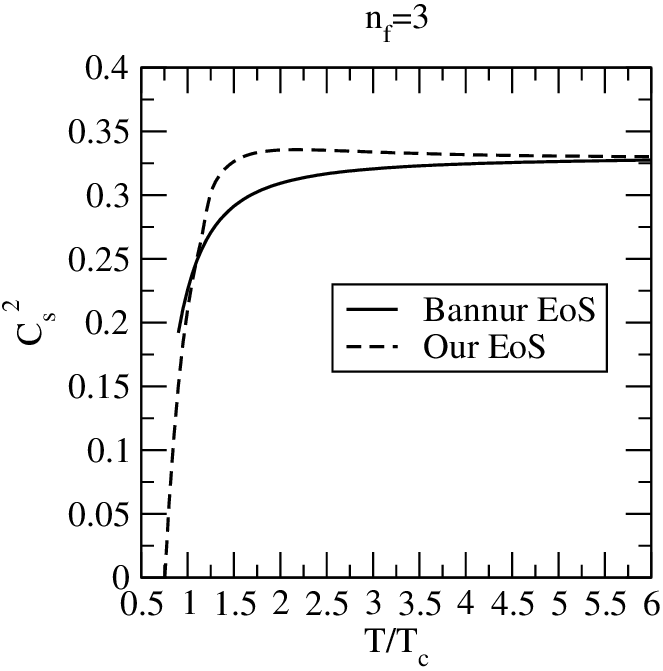}
\hspace{7mm}
\includegraphics[scale=.55]{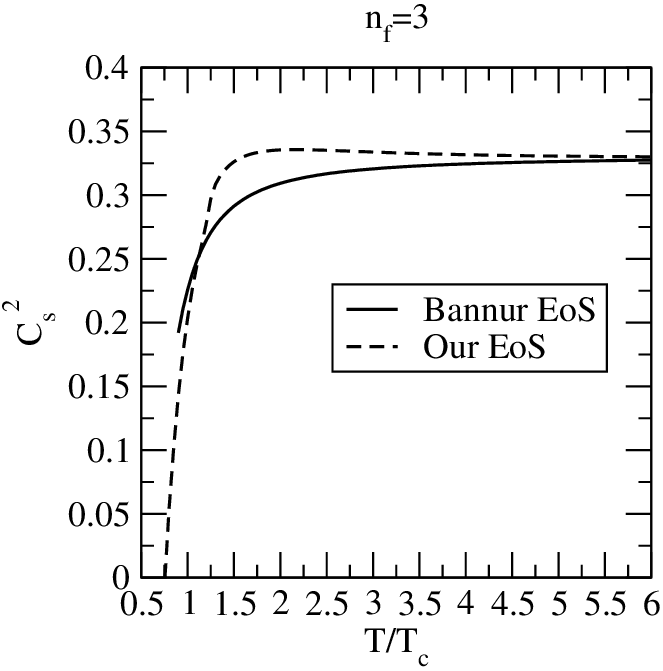}
\vspace{-1mm}
\caption{Plots of $c_s^2$ as a function of $T/T_c$ for Bannur EoS, 
Our EoS (using quasi-particle Debye mass) for 3-flavor QGP (extreme left figure) for EoS1~\cite{zhai,arnold},
and for EOS2~\cite{kajantie} (extreme right figure). The notations are same as Figure2.} 
\vspace{30mm}
\label{cs25}
\end{figure*}
\subsection{Imaginary part of the potential in the isotropic medium}
To obtain the imaginary part of the potential in the QGP medium,  
the temporal component of the symmetric propagator from in the static
 limit has been considered, which reads~\cite{laine,nilima},
\begin{equation}
Im D^{00}_{F(iso)}(0,k)=\frac{-2\pi T m_D^2}{k(k^2+m_D^2)^2}.
\label{isof}
\end{equation}
Now the imaginary part of the dielectric function in the QGP medium as:
\begin{eqnarray}
\frac{1}{\epsilon (k)}= -\pi T m_D^2 \frac{k^2}{k(k^2+m_D^2)^2}.
\end{eqnarray}
Afterwards,  the imaginary part of the in medium potential is easy to obtain owing the definition of the potential~Eq. (\ref{cornell_pot}) as
 done in~\cite{patra_complex}:
\begin{eqnarray}
Im V(r,T)&=&-\int \frac{d^3\mathbf{k}}{(2\pi)^{3/2}}
(e^{i\mathbf{k} \cdot \mathbf{r}}-1)\nonumber\\
&&\times \left(-\sqrt{\frac{2}{\pi}}\frac{\alpha}{k^2}-\frac{4 \sigma}{\sqrt{2 \pi k^4}}\right) \frac{-\pi T m_D^2\ k}{(k^2+m_D^2)^2}\nonumber\\&&
\end{eqnarray}
After performing the integration,  we find 
\begin{eqnarray}
Im V_{(iso)}(\hat{s},T)=T\left(\frac{\alpha {\hat s^2}}{3}
-\frac{\sigma {\hat s}^4}{30m_D^2}\right)\log(\frac{1}{\hat s}).
\label{imis}
\end{eqnarray}
where $(\hat{s}) = r m_D$
\subsection{Effective fugacity quasi-particle model(EQPM)}
In our calculation, we use the Debye mass $m_D$ for full QCD:
\begin{eqnarray}
m^2_D &=& g^2(T) T^2 \bigg[
\bigg(\frac{N_c}{3}\times\frac{6 PolyLog[2,z_g]}{\pi^2}\bigg)\nonumber\\&&
+{\bigg(\frac{N_f}{6}\times\frac{-12 PolyLog[2,-z_q]}
{\pi^2}\bigg)}\bigg].
\end{eqnarray}
Here, $g(T)$ is the QCD running coupling constant, $N_c=3$ ($SU(3)$) and $N_f$
 is the number of flavor, the function $PolyLog[2,z]$ having form,
 $PolyLog[2,z]=\sum_{k=1}^{\infty} \frac{z^k}{k^2}$ and $z_g$ is the quasi-gluon effective fugacity and $z_q$ is quasi-quark
 effective fugacity. These distribution functions are isotropic
 in nature. These fugacities should not be confused with any
 conservations law (number conservation) and have merely been
 introduced to encode all the interaction effects at high temperature
 QCD. Both $z_g$ and $z_q$ have a very complicated temperature
 dependence and asymptotically reach to the ideal value unity
\cite{chandra2}. The temperature dependence $z_g$ and $z_q$  fits well to the 
form given below,
\begin {equation}
\label{eq2}
z_{g,q}=a_{q,g}\exp\bigg(-\frac{b_{g,q}}{x^2}-\frac{c_{g,q}}{x^4} -\frac{d_{g,q}}{x^6}\bigg).
\end {equation}
(Here $x=T/T_c$ and $a$, $b$, $c$ and $d$ are fitting parameters), for both EOS1 and EOS2. Here, EoS1 is the $O(g^5)$ hot
 QCD \cite{zhai,arnold} and EoS2 is the $O(g^6\ln(1/g)$ hot QCD EoS \cite{kajantie} in the quasi-particle description \cite{chandra1,chandra2} respectively. 
 Now, the expressions for the Debye mass can be rewritten in terms of effective charges for the quasi-gluons and quarks as:
 \begin{equation}
m^2_D=
\left\{\begin{array}{rcl}
Q^2_g T^2\frac{N_c}{3} &\mbox{for pure gauge,} &\\
T^2( \frac{N_c}{3} Q^2_g)+(\frac{N_f}{6} Q^2_q) &\mbox{for full QCD}&
\end{array} \right.
\end{equation}
where, $Q_g$ and $Q_q$ are the effective charges given by the equations:
\begin{eqnarray}
 Q^2_g&=&g^2 (T) \frac{6 PolyLog[2,z_g]}{\pi^2}\nonumber\\
 Q^2_q&=&g^2 (T)  \frac{-12 PolyLog[2,-z_q]}{\pi^2}.
\end{eqnarray}
 
In our present analysis we had used the temperature dependence of the quasi-particle Debye mass, $m_D^{QP}$ in full QCD with $N_f= 3$ to determine
charmonium suppression in an expanding, dissipative strongly interacting QGP medium. This quasi-particle Debye mass,
$m_D^{QP}$ has the following form:
\begin{eqnarray}
 m_D^{QP} &=& \frac{2}{\pi^2} g(T) T \bigg[\frac{N_c}{3} PolyLog[2,z_g] \nonumber\\&&-N_f PolyLog[2, -z_q] \bigg]^{\frac{1}{2}}.
\end{eqnarray}
\section{Binding energy and Dissociation Temperature}
To obtain the binding energies with heavy quark potential we need to solve the Shr\"{o}dinger equation numerically. In the limiting case 
discussed earlier, the medium modified  potential resembles to the hydrogen
 atom problem~\cite{Matsui:86}.
 The solution of the Schr\"{o}dinger equation gives the eigenvalues for the 
ground states and the first excited states in charmonium
 ($\jpsi$, $\psi^\prime$ etc.) and bottomonium
 ($\Upsilon$, $\Upsilon'$ etc.) spectra :
\begin{eqnarray}
\rm{Re}~E_{\rm{bin}}^{\rm{iso}}\stackrel{\hat{s}\gg 1}{=}\left( \frac{m_Q\sigma^2 }{m_{{}_D}^4 n^{2}} +\alpha 
m_{{}_D} \right);~n=1,2 \cdot \cdot \cdot
\end{eqnarray}
where $m_Q$ is the mass of the heavy quark. 

In our analysis, we have fixed the critical temperature ($T_c$ = $0.197 GeV$) and have taken the quark masses $m_Q$, as $m_{\Upsilon} = 4.5~$GeV,
$m_{\Upsilon'} = 5.01~$GeV and $m_{\chi_b} = 5.18$GeV, as calculated in ~\cite{Aulchenko:2003qq} and the string tension ($\sigma$) is taken as $0.184GeV^2$. 
Let us now proceed to the computation of the dissociation temperatures for the above mentioned quarkonia bound states.

As we know, dissociation of a quarkonia bound state in a thermal QGP medium will occur whenever the binding energy, $E_B$ of the said state will fall below the 
mean thermal energy of a quasi-parton. In such situations the thermal effect can dissociate the quakonia bound state. To obtain the lower bound of the 
dissociation temperatures of the various quarkonia states, the (relativistic) thermal energy of the partons will $3\ T$. 
 The dissociation is suppose occur whenever,
\begin{eqnarray}
\rm{Re}~E_{\rm{bin}}^{\rm{iso}}\stackrel{\hat{s}\gg 1}{=} E_B (T_D)= 3 T_D.
\end{eqnarray} 

 The $T_D$'s  for the $b\bar{b}$ sates $\Upsilon$, $\Upsilon'$ and $\chi_b$ with the dissociation temperature are listed in Table I and II for
for EoS1 and EoS2 respectively . We observe that (on the basis of temperature dependence of binding energy)
 $\Upsilon'$ dissociates at lower temperatures as compared to $\Upsilon$ and $\chi_b$ for both the equations of state.
 
\begin{table}
\label{table1}
\centering
\caption{Dissociation temperature$T_D$ (for a 3-flavor QGP),
using quasi-particle debye mass for bottomonium states, for EoS1.}
\vspace{3mm}
\begin{tabular}{|l|l|l|l|l|l|l|l|}
\hline
State &$\tau_F$  &$T_D$  &$c_s^2$(SIQGP)  &$c_s^2$(Id)  &$\epsilon_s$(SIQGP)  &$\epsilon_s$(Id) \\
\hline\hline
$\Upsilon$ &0.76&  1.98 & 0.335 &1/3 & 24.39 &23.89 \\
\hline
$\Upsilon'$ & 1.90& 1.53 & 0.326 &1/3 & 8.28 &8.16\\
\hline
$\chi_b$ &2.60& 1.61 & 0.331 &1/3 & 10.21 &10.10\\
\hline
\end{tabular}
\end{table}

\begin{table}
\label{table4}
\centering
\caption{Dissociation temperature$T_D$ (for a 3-flavor QGP),
using quasi-particle debye mass for bottomonium states, for EoS2.}
\vspace{3mm}
\begin{tabular}{|l|l|l|l|l|l|l|l|}
\hline
State &$\tau_F$  &$T_D$  &$c_s^2$(SIQGP)  &$c_s^2$(Id)  &$\epsilon_s$(SIQGP)  &$\epsilon_s$(Id) \\
\hline\hline
$\Upsilon$ &0.76&  2.04 & 0.335 &1/3 & 27.05 &27.09 \\
\hline
$\Upsilon'$ & 1.90& 1.58 & 0.328 &1/3 & 9.35 &9.44\\
\hline
$\chi_b$ &2.60& 1.65 & 0.331 &1/3 & 11.21 &11.34\\
\hline
\end{tabular}
\end{table}
\section{Formulation}
In relativistic nucleus-nucleus collisions the equation of state for the quark matter is an important observable and
the properties of the matter are sensitive to it. The expansion of QGP is
quite sensitive to EoS through the speed of sound,explores the sensitivity of the quarkonium suppression to the equation of state~\cite{dpal,compare}. 

For a strongly-coupled QGP Bannur~\cite{banscqgp} developed an equation of state by incorporating running coupling constant
and did a appropriate modifications to take account color and flavor degrees of freedom and obtained a reasonably good fit to the lattice results.
Now we will discuss briefly the equation of state which is expressed as a function of plasma parameter
$\Gamma$~\cite{ic.1}:
\begin{equation} 
\epsilon_{{}_{\rm{QED}}} = \left( \frac{3}{2} + u_{ex} (\Gamma) 
\right) \, n \, T \; , \label{eq:scp} 
\end{equation}
Plasma parameter $\Gamma$, is the ratio of average potential 
energy to average kinetic energy of particles, is assumed to be weak 
($<<1$) and is given by:
\begin{equation} \Gamma \equiv \frac{<PE>}{<KE>} = \frac{Re[V({\bf r},T)]}{T} \; ,\end{equation}
We have studied the variation of plasma parameter with temperature and as well
with the number of flavours present in the system and are shown in Fig. \ref{gama1} for EoS1 and EoS2 respectively. As
the temperature increases, potential becomes weaker and hence the plasma parameter have started waning, albeit 
at very large temperature it increases slightly due to the contribution coming from the (positive)
finite-range terms in the potential, unlike the decreasing trend in Bannur model\cite{banscqgp} always due to the presence of Coulomb interaction alone
in the deconfined phase.  

Let us consider that hadron exists for $T<T_c$ and goes to QGP for $T>T_c$ for strongly-coupled plasma in QCD.
As it was assumed that confinement interactions due to QCD vacuum 
has been melted~\cite{banscqgp} at $T=T_c$ and thus for $T>T_c$, it is the strongly interacting 
plasma of quarks and gluons and no glue balls or hadrons .
After inclusion of relativistic and quantum effects, 
the equation of state which has been obtained in the plasma parameter can be written as: 
\begin{equation} 
\varepsilon = \left( \frac{}{} 3 + u_{ex} (\Gamma) \right) 
\, n \, T \;, \end{equation}
Now, the scaled-energy density is written as in terms of ideal contribution
\begin{equation} 
e(\Gamma) \equiv \frac{\varepsilon}{\varepsilon_{SB}} 
= 1 + \frac{1}{3} u_{ex} (\Gamma) \quad,
\end{equation}
where $\varepsilon_{{}_{SB}}$ is given by,
\beq
\varepsilon_{{}_{SB}} \equiv (16 + 21 \,n_f /2)\pi^2 T^4 /30, 
\eeq 
Here, $n_f$ is the number of flavor of quarks and gluons. 
Now, we will employ two-loop level QCD running coupling constant in $\msbar$ scheme~\cite{shro},:
\bqa
\label{eqg}
{g^2}(T) \approx 2 b_0 \ln \frac{\bar\mu}{\lambdamsbar}
{\left( 1 + \frac{b_1}{2b_0^2} \frac{\ln \left( 2 \ln 
\frac{\bar \mu}{\lambdamsbar}\right)}{\ln \frac{\bar\mu}{\lambdamsbar}}
\right)}^{-1} \quad ,
\eqa
Here $b_0= (33-2n_f)/(48 \pi^2)$ and $b_1= (153-19n_f)/(384\pi^4)$. In $\msbar$ scheme,
$\lambdamsbar$ and $\bar \mu$ are the renormalization scale and the scale parameter  respectively. 
For, the EoS to depend on the renormalization scale,
the physical observables should be scale independent. 
We invade the problem by trading off the dependence on renormalization scale ($\lambdamsbar$)
to a dependence on the critical temperature $T_c$.
\bqa
\bar\mu~\exp(\gamma_E+c)&=&\lambdamsbar (T)\nonumber\\
\lambdamsbar (T) \exp(\gamma_E+c)&=&4 \pi \Lambda_T \quad,
\eqa
here $\gamma_E$=0.5772156 and $c=\left ( n_c-4n_f \ln 4 \right)/\left(22n_c-n_f \right)$,
which is a constant depending on colors and flavors.
There are several incertitude, associated 
with the the scale parameter $\bar \mu$ and renormalization scale $\lambdamsbar$,  which occurs in the
expression used for the running coupling constant $\alpha_s$. This issue
has been considered well in literature and resolved by the BLM criterion due to Brodsky, Lepage 
and Mackenzie~\cite{shaung}. $\lambdamsbar$ is allowed to vary between $\pi T$ and $4 \pi T$~\cite{braten} .
For our motive, we choose the $\lambdamsbar$ close to the central value $2 \pi T_c$~\cite{vuorinen} for $n_f$=0 and 
 for both $n_f$=2 and $n_f$=3 flavors the value is $\pi T_c$. If the factor
 $\frac{b_1}{2b_0^2} \frac{\ln \left( 2 \ln \frac{\bar \mu}{\lambdamsbar}\right)}{\ln \frac{\bar\mu}{\lambdamsbar}}$
is $ \ll 1$ then the above expression reduces to the expression used in~\cite[Eq.(10)]{banscqgp}, after neglecting the higher order
terms of the above factor.
However, this possibility does not hold good for the temperature ranges used in the calculation and 
cause an error in coupling which finally makes the difference in the results between our model and Bannur model~\cite{banscqgp}. 
first of all, we will calculate the energy density $\varepsilon (T)$ from Eq.(25) and using the thermodynamic relation, 
\begin{equation} 
\varepsilon = T \frac{dp}{dT} - P \quad, 
\end{equation}
we calculated the pressure as 
\begin{equation} \frac{P}{T^4} = \left( \frac{P_0}{T_0} + 3 a_f \int_{T_0}^T \, 
d\tau \tau^2 e(\Gamma(\tau)) \right) / T^3 \; , \label{eq:p} \end{equation} 
here $P_0$ is the pressure at some reference temperature $T_0$. Now, the speed of sound $c_s^2 (= \frac{dP}{d\varepsilon})$ can be 
calculated once we know the pressure $P$ and energy density $\varepsilon$. 

\begin{figure*}
\vspace{-2mm}
\includegraphics[scale=.50]{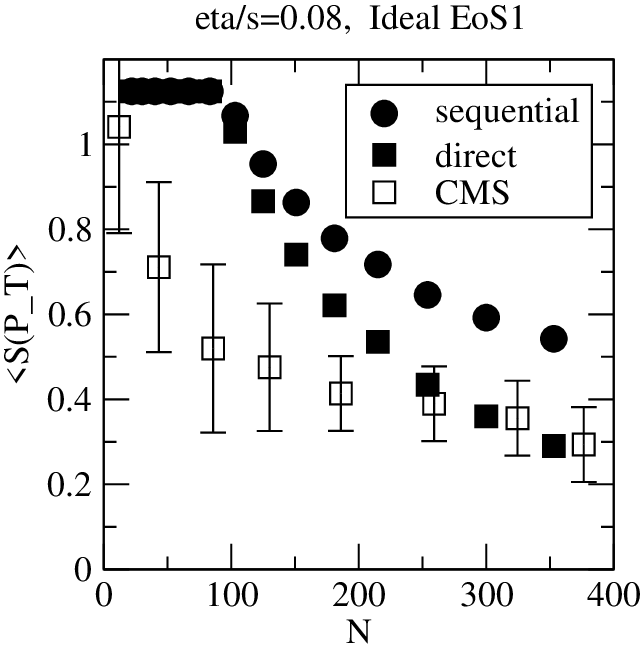}
\hspace{6mm}
\includegraphics[scale=.50]{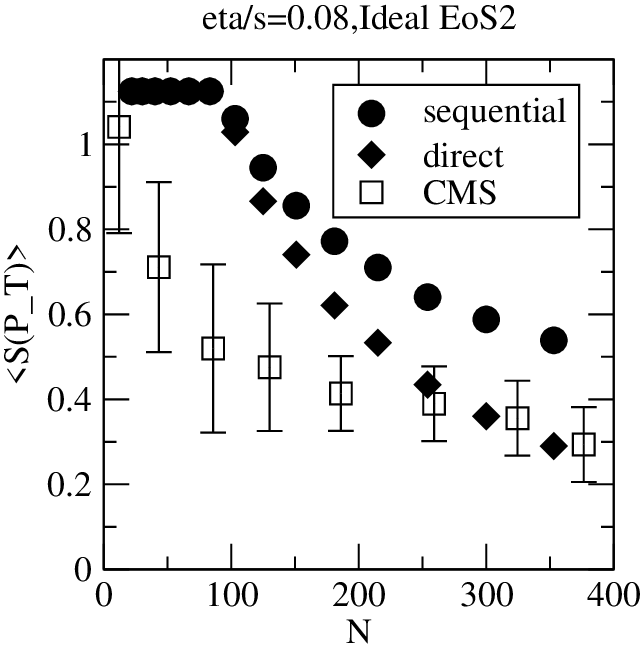}
\vspace{-1mm}
\caption{The variation of $\pt$ integrated survival probability versus $N$ for $\Upsilon$ at $\sqrt{S_{NN}}$= 2.76 TeV with preliminary CMS data~\cite{cms}. 
The experimental data are shown by the squares with error bars whereas circles and diamond represent with ($\sincl$) without ($\sdir$) sequential 
melting using the value of $T_D$'s and related parameters from Table I and Table II for Ideal equation of state. Left panel shows EoS1 and right panel shows EoS2 .}
\vspace{37mm}
\end{figure*}
\begin{figure*}
\vspace{-2mm}
\includegraphics[scale=.50]{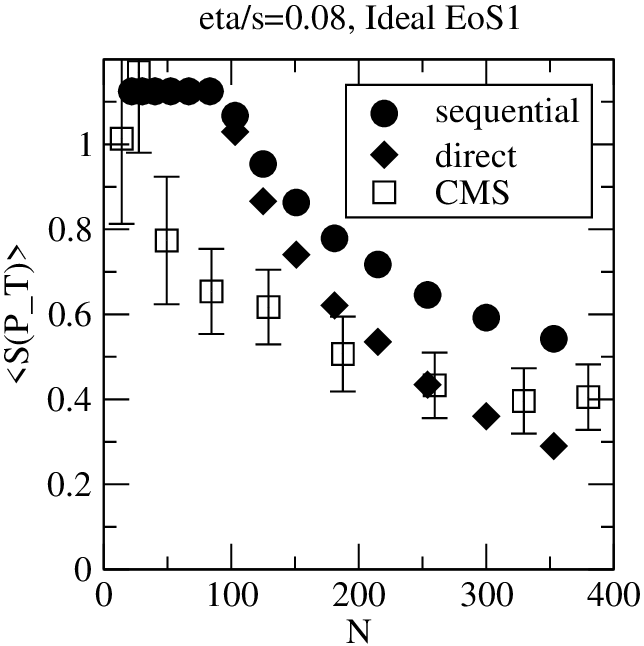}
\hspace{6mm}
\includegraphics[scale=.50]{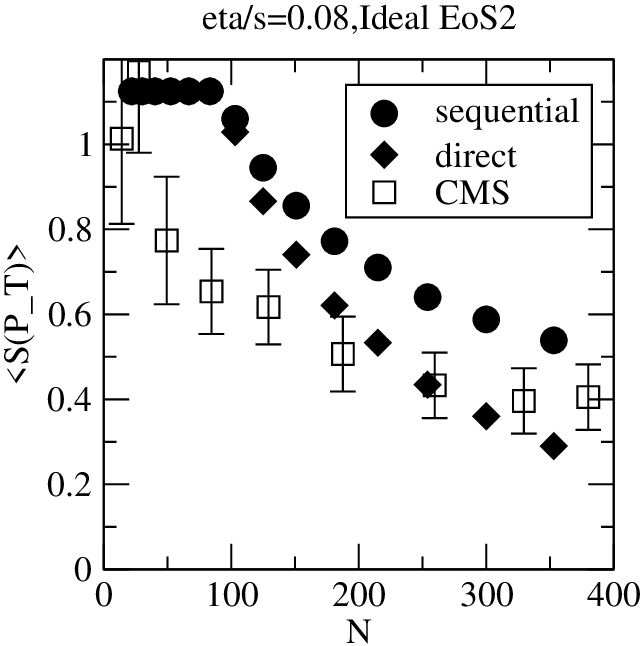}
\vspace{-1mm}
\caption{Same as Fig.5 but the variation of $\pt$ integrated survival probability versus $N$ for $\Upsilon$ at $\sqrt{S_{NN}}$= 5.02 TeV with preliminary 
CMS data ~\cite{chad}.}
\vspace{35mm}
\end{figure*}

\begin{figure*}
\vspace{-2mm}
\includegraphics[scale=.50]{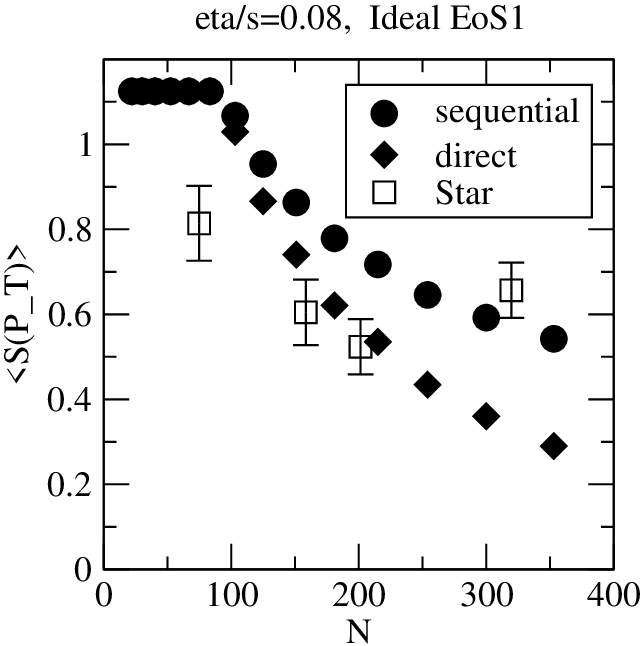}
\hspace{6mm}
\includegraphics[scale=.50]{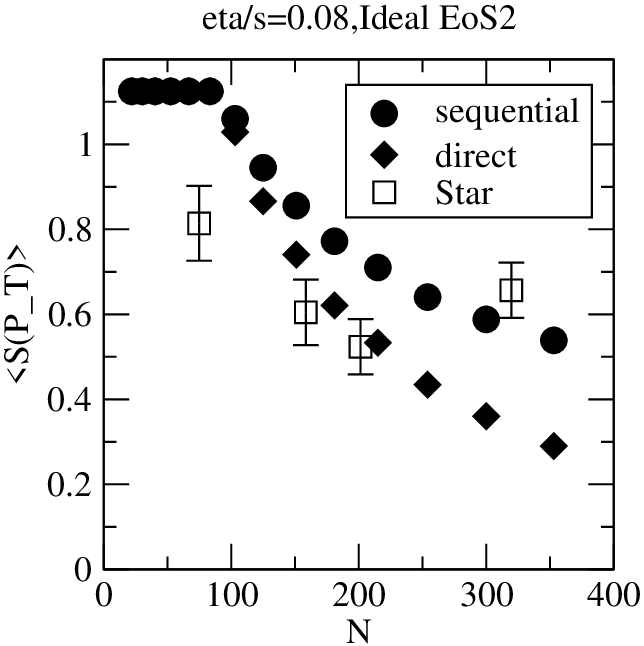}
\vspace{-1mm}
\caption{Same as Fig.5 but the variation of $\pt$ integrated survival probability versus $N$ for $\Upsilon$ at $\sqrt{S_{NN}}$= 200 GeV
with preliminary STAR data~\cite{ye}.}
\vspace{35mm}
\end{figure*}

\begin{figure*}
\vspace{-2mm}
\includegraphics[scale=.50]{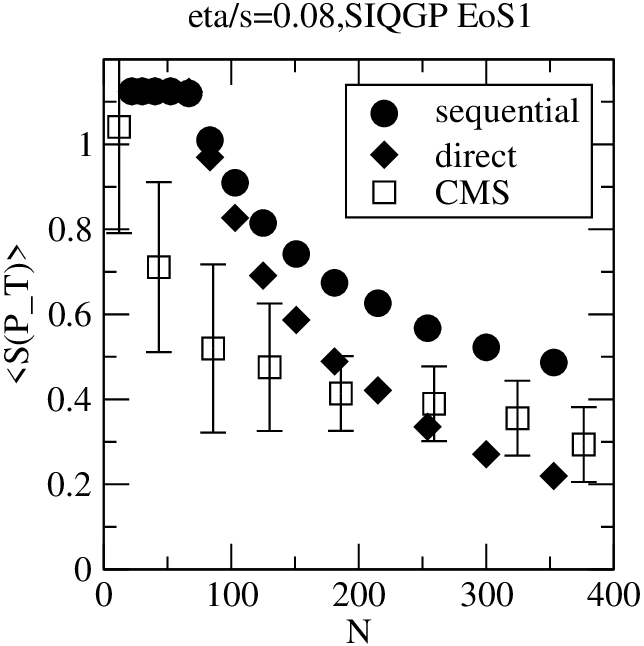}
\hspace{6mm}
\includegraphics[scale=.50]{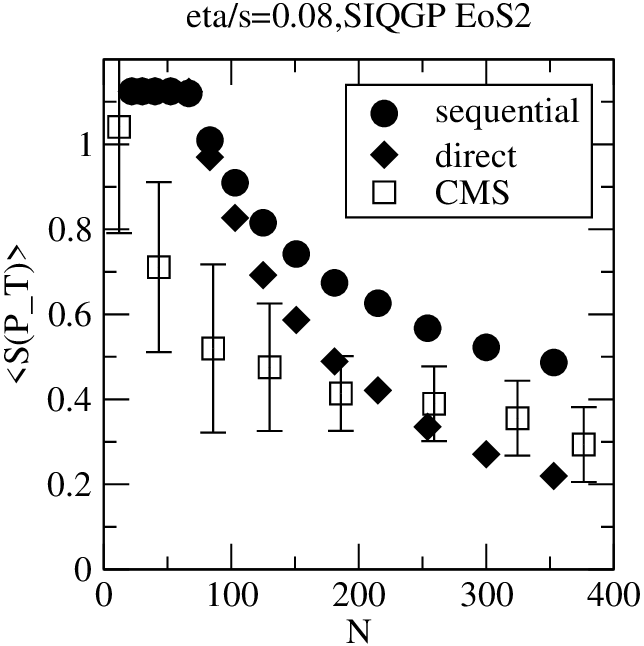}
\vspace{-1mm}
\caption{The variation of $\pt$ integrated survival probability versus $N$ for $\Upsilon$ at $\sqrt{S_{NN}}$= 2.76 TeV with preliminary CMS data~\cite{cms}.
The experimental data are shown by the squares with error bars whereas circles and diamond represent with ($\sincl$) without ($\sdir$) sequential melting
using the value of $T_D$'s and related parameters from Table I and Table II for SIQGP equation of state. 
Left panel shows EoS1 and right panel shows EoS2.}
\vspace{35mm}
\end{figure*}

\begin{figure*}
\vspace{-2mm}
\includegraphics[scale=.50]{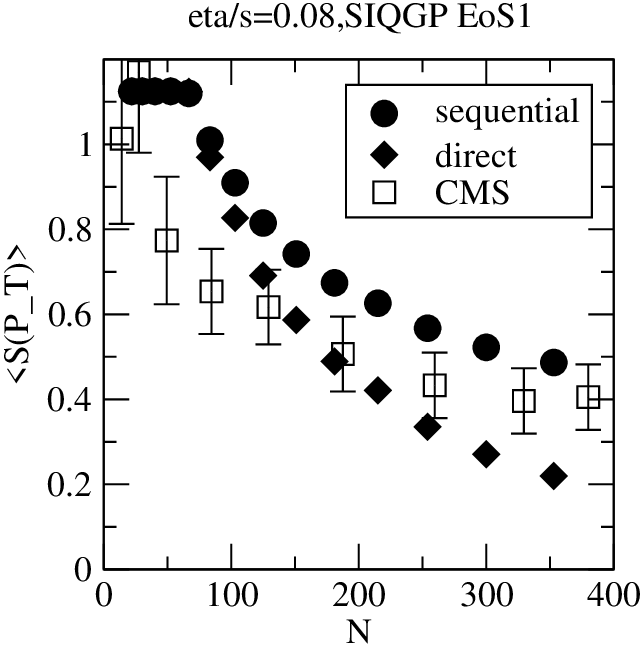}
\hspace{6mm}
\includegraphics[scale=.50]{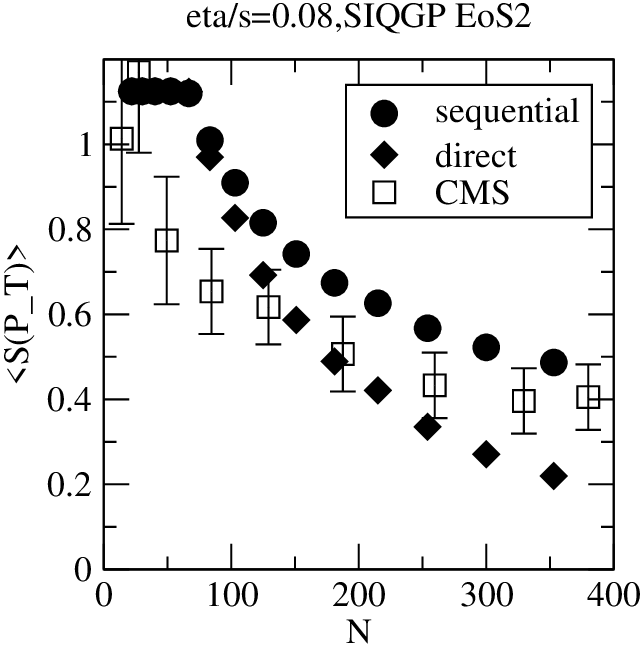}
\vspace{-1mm}
\caption{Same as Fig.8 but the variation of $\pt$ integrated survival probability versus $N$ for $\Upsilon$ at $\sqrt{S_{NN}}$= 5.02 TeV with
preliminary CMS data~\cite{chad}.}
\vspace{35mm}
\end{figure*}

\begin{figure*}
\vspace{-2mm}
\includegraphics[scale=.50]{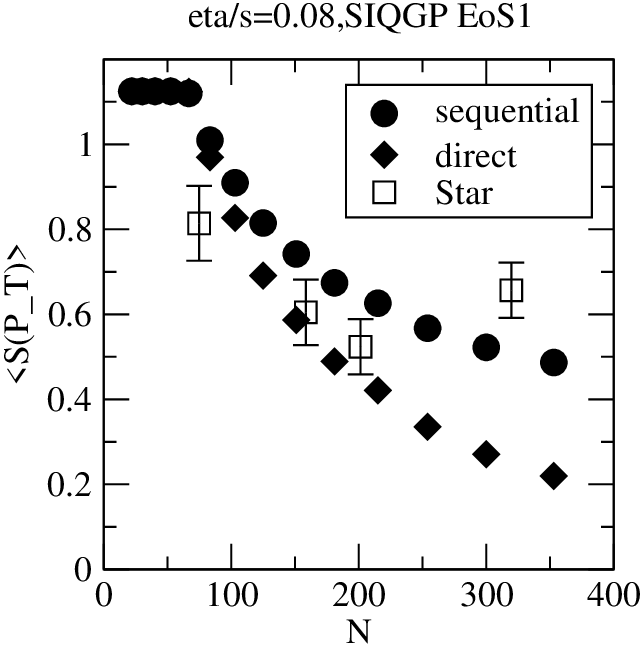}
\hspace{6mm}
\includegraphics[scale=.50]{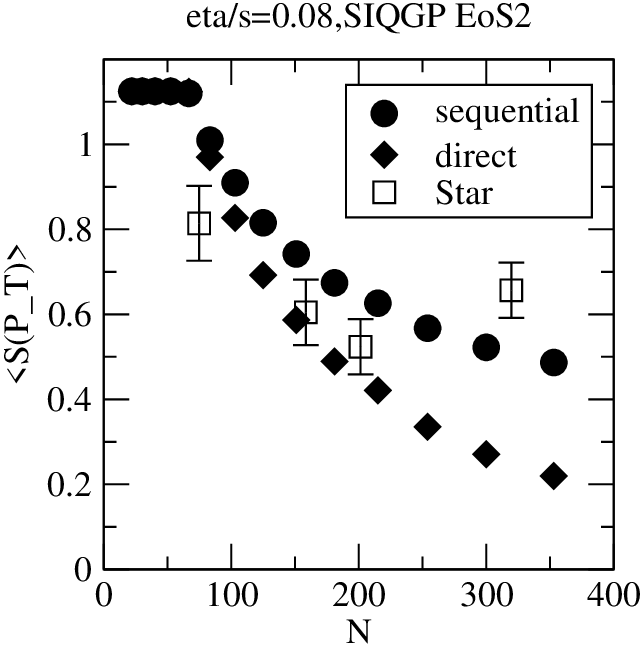}
\vspace{-1mm}
\caption{Same as Fig.8 but the variation of $\pt$ integrated survival probability versus $N$ for $\Upsilon$ at $\sqrt{S_{NN}}$= 200 GeV with
preliminary STAR data~\cite{ye}.}
\vspace{35mm}
\end{figure*}
\section{Survival of bottomonium state}
In order to derive the $\Upsilon$ survival probability for an expanding QGP firstly, we explore the effects of dissipative terms up to first-order in the stress-tensor. 
In the presence of viscous forces, the energy-momentum tensor is written as,
\begin{equation}
T^{\mu\nu}-\pi^{\mu\nu}= (\epsilon+p)u^\mu u^\nu + g^{\mu \nu} p ,
\label{tmun}
\end{equation}
where the stress-energy tensor, $\pi^{\mu\nu}$ up to first-order 
is given by
\beq
\pi^{\mu\nu}=\eta \langle \nabla^\mu u^\nu \rangle ~, 
\eeq
where $\eta$ is the co-efficient of the shear viscosity and 
$ \langle \nabla^\mu u^\nu \rangle$ is the symmetrized velocity gradient. 

In Bjorken expansion, the equation of motion is given by
\beq
\partial_\tau \epsilon+\frac{\epsilon+p}{\tau} =\frac{4 \eta}
{3 \tau^2}\, .
\label{3.16}
\eeq
The solution of equation of motion  (\ref{3.16}) is given as,
\begin{eqnarray}
\label{eqs1}
\epsilon(\tau) \tau^{(1+c_s^2)}+ \frac{4a}{3{\tilde{\tau}}^2}
\tau^{(1+c_s^2)}
&=&\epsilon(\tau_i)\tau_i^{(1+c_s^2)}
+\frac{4a}{3 {\tilde{\tau_i}}^2} \nonumber\\
&=&{{\mbox{const}}}~,
\end{eqnarray}

where the constant
\beq
a = {\left(\frac{\eta}{s}\right)T^3_i \tau_i} 
\eeq
 and the symbols,
\beq
{\tilde{\tau}}^2 = (1-c_s^2)\tau^2
\eeq
and 
\beq
{\tilde{\tau}}_i^2 = (1-c_s^2)\tau_i^2.
\eeq
The first term accounts for the contributions coming from the 
zeroth-order expansion (ideal fluid) and the second term is the first-order 
viscous corrections.
We now have all the ingredients to write down the survival probability.
Chu and Matsui~\cite{chu} studied 
the transverse momentum dependence ($p_T$) of the survival probability
by choosing the speed of sound $c_s^2=1/3$ (ideal EoS)
and the extreme value $c_s^2=0$.
Instead of taking arbitrary values of $c_s^2$ we tabulated
the values of $c_s^2$ in Tables I and II corresponding to the
dissociation temperatures for bottomonium states for EOS1 and EOS2.
One can define initial energy density $\epsilon_i$ as
\begin{equation}
\label{eps}
\epsilon_i=(1+\beta)  \langle \epsilon_i \rangle ~~~;\beta=1.
\end{equation} 
Here, $\beta$ represents the proportionality of the deposited energy to the nuclear thickness whearas ${\langle \epsilon_i \rangle}$ is the average initial energy density 
and will be given by the modified Bjorken formula~\cite{NA50,khar}:
\begin{equation}
\label{eps1}
{\langle \epsilon_i \rangle}=\frac{\xi}{A_T\,\tau_i}
\left(\frac{dE_T}{dy_h}\right)_{y_h=0} ,
\end{equation}
where $A_T$ is the transverse overlap area of the colliding nuclei and 
$(dE_T/dy_h)_{y_h=0}$ is the transverse energy deposited per unit 
rapidity. we use the experimental value of the transverse overlap area
$A_T$ and the pseudo-rapidity distribution ${dE_T/d\eta_h \mid}_{\eta_h=0}$
~\cite{sscd} at various values of  number of participants
$N_{part}$. These ${dE_T/d\eta_h \mid}_{\eta_h=0}$ numbers are then multiplied
by a Jacobian 1.25 to yield the rapidity distribution
${dE_T/dy_h \mid}_{y_h=0}$ which will be further used to calculate
the average initial energy density from Bjorken formula (\ref{eps1}). After getting the value of average initial energy density we 
can obtained the initial energy density from the formula (\ref{eps}). The scaling factor $\xi=5$ has been introduced in order to obtain the 
desired values of initial energy densities  
~\cite{hiran} for most central collision
which are consistent with the predictions of the self-screened parton 
cascade model~\cite{eskola} and also with the 
requirements of hydrodynamic simulation~\cite{hiran} to fit
the pseudo-rapidity distribution of charged particle 
multiplicity $dN_{ch}/d\eta$ for various centralities 
observed in PHENIX experiments at RHIC energy. 
Let $\phi$ is the angle between the transverse momentum and position vector $r_{\Upsilon}$. Now assuming that $b\bar b$ is formed inside
screening region at a point whose position vector is $\vec r$ and moves with transverse momentum $p_{T}$ making an azimuthal angle. 
Then the condition for escape of $b\bar b$ without forming bottomonium states is expressed as: 
\begin{equation}
\cos\phi \geq Y; \; \; Y = \frac{(r_{s}^{2} - r_{\Upsilon}^{2})m -
\tau_{F}^{2}p_{T}^{2}/m}{2r_{\Upsilon}\tau_{F}p_{T}},
\label{phi}
\end{equation}
where, ${r_\Upsilon}$ is the position vector at which the bottom, anti bottom quark pair is
formed, $\tau_{F}$ is the proper formation time required for the formation of
bound states of $b\bar{b}$ from correlated $b \bar b$ pair and $m$ is
the mass of bottomonia ($m = M_{\Upsilon},\;\; M_{\chi_{b}},\;\; M_{\Upsilon'}$ for different resonance states of bottomonium).
Assuming the radial probability distribution for the production of $b\bar b$ pair in hard collisions at transverse distance $r$ as 
\begin{equation}
f(r)\propto\left(1-\frac{r^2}{R_T^2}\right)^{\alpha}\theta(R_T-r).
\end{equation}    
Here we take $\alpha=0.5$ in our calculation as used in Ref.~\cite{chu}. Then, in the colour screening scenario, the survival probability
for the bottomonium in QGP medium can be expressed as~\cite{mmish,chu} :
\begin{equation}
S(p_T,N_{part})=\frac{2(\alpha+1)}{\pi R_T^2}\int_0^{R_T}dr r \phi_{max}(r)\left\{1-\frac{r^2}{R_T^2}\right\}^{\alpha},
\end{equation}
where the maximum positive angle $\phi_{max}$ allowed by Equation 26 becomes~\cite{suppr} :
$$
\phi_{max}(r)=\left\{\begin{array}{rl}
\pi     & \mbox{~~if $Y\le -1$}\\
\pi-\cos^{-1}|Y|  & \mbox{~~if $0\ge Y\ge -1$}\\
\cos^{-1}|Y| & \mbox{~~$0\le Y\le -1$}\\
 0        & \mbox{~~$Y \ge 1$} 
\end{array}\right.
$$
since the experimentalists always measure the quantity namely $p_T$ integrated nuclear modification factor. We get the theoretical 
$p_T$ integrated survival probability as follows :

\begin{equation}
S (N_{part})=\frac{\int_{p_{Tmin}}^{p_{Tmax}}S(p_T,N_{part})dp_T}{\int_{p_{Tmin}}^{p_{Tmax}}d p_T}.
\end{equation} 

In nucleus-nucleus collisions, it is known that only about
60\% of the observed $\Upsilon$ originate directly in hard collisions while
30\% of them come from the decay of $\chi_b$
and 10\% from the decay of $\Upsilon'$. Hence, the $\pt$-integrated inclusive
survival probability of $\Upsilon$ in the QGP becomes~\cite{satz,dpal}.
\begin{equation}
\langle S^{{}^{\rm incl}} \rangle = 0.6 {\sdir}_{{}_\Upsilon}
+0.3 {\sdir}_{{}_{\chi_b}}
+0.1 {\sdir}_{{}_{\Upsilon'}}
\end{equation}

\section{Results and discussions}
In our results we had obtained the 
variation of plasma parameter with temperature and as well with the number of flavours present in the system
and are shown in Fig. \ref{gama1} for EoS1 and EoS2 respectively. After then,
 in Fig. \ref{pres5}, we have plotted the variation of pressure ($P/T^4$) with
temperature ($T/T_c$) using EoS1 and EoS2 for 3-flavor QGP along with Bannur EoS~\cite{banscqgp} and compared it with lattice results~\cite{banscqgp, boyd}.
For each flavor, $g_c$ and $\Lambda_T$ are adjusted to get a good fit to lattice results in Bannur Model.
Now, energy density $\varepsilon$, speed of sound $c_s^2$ etc. can be derived
since we had obtained the pressure, $P(T)$ .
In Fig. \ref{eps2}, we had plotted the energy density 
($\varepsilon / T^4$) with temperature ($T/ T_c$) using EoS1~\cite{zhai,arnold} and EoS2 for 3-flavor QGP along with Bannur EoS~\cite{banscqgp}
and compared it with lattice result~\cite{banscqgp, boyd}. 
As the flavor increases, the curves shifts to left.
In Fig. \ref{cs25}, the speed of sound, $c_s^2$ is plotted for all three systems, using EoS1 and EoS2 for 3-flavor QGP along 
with Bannur EoS~\cite{banscqgp}. Since lattice results are not available for 3-flavor, therefore comparison
has not been checked for the above mentioned flavor. Our flavored results matches excellent with the lattice results. We observe that 
 as the flavor increases $c_s^2$ becomes larger for both EoS1 and EoS2. All three curves shows similar behaviour, i.e, sharp rise near $T_c$ and then flatten to the
ideal value ($1/3$).

In this paper we had calculated the dissociation temperatures for the bottomonium states 
($\Upsilon$, $\Upsilon'$, $\chi_b$,etc.), by modifiying the Cornell potential and incorporating the quasi-particle debye mass. On that dissociation temperature we had
calculated the screening energy densities, $\eps$ and the speed of sound $c_s^2$ which are also listed in the table I and II for both EoS1 and EoS2 respectively.
We observe from the table I- II that the value of $\epsilon_s$ is different for different bottomonium 
states and varies from one EoS to other. 
If $\eps \gtrsim \epsilon_i$, initial energy density, then 
there will be no suppression at all i.e., survival probability, $\spt$ is equal to 1.
With this physical understanding we analyze our results, $\sinpt$ as a function of the number of participants $N_{{Part}}$ in an expanding QGP. 

Here we are using the values as inputs listed in Table I and Table II, to calculate $\sinpt$ for both EOS1 and EOS2 respectively.
The experimental data (the nuclear-modification factor $R_{AA}$) are shown by the squares with error bars whereas
circles represent sequential suppression. We had compared our results with the experimental results for the case of $\eta/s=0.08$ for both EoS1 and EoS2 and found good
agreement. We observe from the figs. 5-10 that $\sinpt$ for both the directly and sequentially produced Upsilon ($\Upsilon$) are quite high with the higher values
of $T_D$'s which is obtained from EOS2 (in Table II) compared to EOS1 (in Table I) for both SIQGP and Ideal equation of states. We find that the survival probability of
sequentially produced $\Upsilon$ is slightly higher compared to the directly produced $\Upsilon$ and is closer to the experimental results. We also observed that sequentially 
produced $\Upsilon$ nicely matches for the EOS1 compared to the EOS2.
The smaller value of screening energy density $\epsilon_s$ causes an increase in the screening time and results in more suppression to match with the experimental results.

\section{Conclusions}
We studied the equation of state for strongly interacting quark-gluon plasma in the framework of strongly coupled plasma 
with appropriate modifications to take account of color and flavor degrees of freedom and QCD running coupling constant.
In addition, we incorporate the nonperturbative effects in terms of nonzero string tension in the 
deconfined phase, unlike the Coulomb interactions alone in the deconfined phase beyond the critical temperature.
Our results on thermodynamic observables {\em viz.} pressure, energy density, speed of sound etc. nicely fit the
results of lattice equation of state.
We had then calculated the dissociation temperatures for the bottomonium states 
($\Upsilon$, $\Upsilon'$, $\chi_b$,etc.), by incorporating the quasi-particle debye mass. On that dissociation temperature we had
calculated the screening energy densities, $\eps$ and the speed of sound $c_s^2$ which are listed in the table I and II for both EoS1 and EoS2 respectively.
 By using the above quantities as a input we have then studied
the sequential suppression for bottomonium states at the LHC energy in a longitudinally expanding partonic system, which underwent through the
successive pre-equilibrium and equilibrium phases in the presence of dissipative forces. Bottomonium suppression in nucleus-nucleus collisions
compared to $p$-$p$ collisions couples the in-medium properties of the bottomonia states with 
the dynamics of the expanding medium. We have found a good agreement with the
 experimental data from RHIC 200GeV/nucleon Au-Au collisions, LHC 2.76 TeV/nucleon Pb-Pb, and LHC 5.02 TeV/nucleon Pb-Pb collisions~\cite{brandon,ryblewski}.
Here our attempt is to understand $\Upsilon$ suppression systematically in SIQGP in anisotropic medium. It would be of interest to extend the present study by incorporating
the contributions of the bulk viscosity. These issues will be taken up separetely in the near future.

\section{Acknowledgement}
VKA acknowledge the UGC-BSR research start up grant No. F.30-14/2014 (BSR)
New Delhi. We record our sincere gratitude to the people of India for their generous support for the research in basic sciences.

\end{document}